# The Frozen Core Approximation and Nuclear Screening Effects in Single Electron Capture Collisions

A. L. Harris

Physics Department, Illinois State University, Normal, IL, USA 61790

**Abstract**

Fully Differential Cross Sections (FDCS) for single electron capture from helium by heavy ion impact are calculated using a frozen core 3-Body model and an active electron 4-Body model within the first Born approximation. FDCS are presented for $H^+$, $He^{2+}$, $Li^{3+}$, and $C^{6+}$ projectiles with velocities of 100 keV/amu, 1 MeV/amu, and 10 MeV/amu. In general, the FDCS from the two models are found to differ by about one order of magnitude with the active electron 4-Body model showing better agreement with experiment. Comparison of the models reveals two possible sources of the magnitude difference: the inactive electron's change of state and the projectile-target Coulomb interaction used in the different models. Detailed analysis indicates that the uncaptured electron's change of state can safely be neglected in the frozen core approximation, but that care must be used in modeling the projectile-target interaction.

## 1. Introduction

The study of electron capture (or charge transfer) collisions serves researchers in both fundamental and applied fields. On the applied side, electron capture collisions are needed in areas as diverse as plasma physics, astrophysics, and biophysics. Modelers, theorists, and experimentalists in these fields rely on accurate collision cross sections in order to understand phenomena and develop new technologies. Without an accurate knowledge of the most fundamental atomic interactions, more sophisticated systems cannot be adequately understood, and this is where a connection between fundamental and applied physics is needed. From a fundamental standpoint, electron capture collisions can provide valuable information about atomic structure and few-body interactions on the atomic scale. They have been used to study effects such as electron correlation, multi-step processes, and nuclear-nuclear Coulomb effects.

While these collision systems have been studied extensively for the last century, there are still many theoretical, computational, and experimental challenges. In particular, the small projectile scattering angles involved in heavy particle collisions make highly detailed measurements difficult, and only in the last two decades have a significant number of differential cross section measurements become available [1–7]. The growing body of highly accurate experimental data has spurred the development and application of numerous theoretical models for the study of the electron capture process. While theory's ability to accurately predict experiment has improved, there are still many remaining challenges. For example, an accurate theoretical treatment of heavy particle continuum wave functions remains quite difficult. Additionally, time-dependent and non-perturbative models often perform better than traditional Born-type models, but they often rely on classical, semiclassical, or other approximations, and can require significant computational resources [8–13].

A frequently used approximation for single electron capture collisions in both perturbative and non-perturbative models is the frozen core approximation in which the uncaptured, inactive electron is assumed to not change state. In the case of a bare particle colliding with a helium atom, the implementation of the frozen core approximation effectively reduces the calculation from a 4-body problem to a 3-body problem, which greatly improves computational runtimes. The frozen core approximation is also often applied to single ionization and single excitation collisions in which there is an inactive electron that can be modeled as a bystander with only screening effects and no change of state. Naturally, a complete 4-body calculation in all of these collision processes represents a more physically accurate description, but the key question is what, if any, limitations exist by using the frozen core approximation.

Thanks to computational advances, there are now many 4-body models in use for the single capture process [2,14–24] with varying degrees of agreement between each other and experiment. Combined with the numerous 3-body models present in the literature, there is no shortage of calculations for the single capture process. However, a focused analysis of the effects of the frozen core approximation in single capture collisions has not been performed. This is the goal of the present investigation.

Previously, we have examined the frozen core approximation in electron and heavy ion impact single ionization of helium [25,26], as well as the 5-body process of $He^+$ + He electron capture [27]. In our study of single ionization, we found that the initial state projectile-target interaction and the final state ionic potential were most influential on the magnitude and shape of the fully differential cross sections (FDCS). The different treatments of these interactions in the 3-Body and 4-Body models represent different approximations for screening of the target nucleus and showed clear effects for both electron and heavy ion projectiles.

For the 5-body $He^+$ + He single capture collision, our analysis also showed that the model used for screening either the projectile or the target nucleus could significantly affect the shape and magnitude of the FDCS. In particular, much like the single ionization case, changes to the initial state projectile-target Coulomb interaction were the primary cause of the differences between the frozen core and active electron FDCS. Based on the consistency of these prior studies, we expect the initial state projectile-target interaction to play an important role in the single capture process, as well. In order to focus solely on the effect of the frozen core approximation, we perform calculations with the first Born approximation (FBA), where the effect of the uncaptured electron can be studied in a straightforward manner. Atomic units are used throughout unless otherwise noted.

## 2. Theory

In both the 3-Body and 4-Body models, the FDCS are calculated using Jacobi coordinates in the center of mass frame and then converted to the lab frame. The fully differential cross section in the center of mass frame is given by [28]

$$\frac{d\sigma^C}{d\Omega} = \frac{(2\pi)^4 \mu_{pa}\mu_{pi}k_f}{k_i}|T_{fi}|^2. \tag{1}$$

where $\mu_{pi}$ is the reduced mass of the scattered projectile and residual ion, $\mu_{pa}$ is the reduced mass of the initial state projectile and target atom, $\vec{k}_f(\vec{k}_i)$ is the center of mass momentum of the scattered (incident) projectile, and $T_{fi}$ is the transition matrix. The differential cross section in the lab frame is related to that of the center of mass frame by

$$\frac{d\sigma^L}{d\Omega} = \left[\frac{(1+2\delta\cos\theta_C+\delta^2)^{\frac{3}{2}}}{|1+\delta\cos\theta_C|}\right]\frac{d\sigma^C}{d\Omega}. \tag{2}$$

where $\delta$ is the ratio of the speed of the center of mass of the entire collision system in the lab frame $V$ and the speed of the scattered projectile in the center of mass frame $v_f^C$, such that $\delta = \frac{V}{v_f^C}$. The angle $\theta_C$ is the scattering angle of the projectile in the center of mass frame, which is related to the lab frame scattering angle by

$$\tan\theta_L = \frac{\sin\theta_C}{\cos\theta_C+\delta}. \tag{3}$$

In order to focus on the effects of the frozen core approximation, we use the first Born approximation (FBA) to calculate the transition matrix for single electron capture from a helium target by bare heavy ion impact. We assume an independent electron model for the target atom and use an uncorrelated target helium wave function. It has been shown that for the single capture process, target electron correlation is unimportant and can be neglected [8]. The incident projectile has charge $Z_p$ and mass $m_p$, and we present results here for H$^+$, He$^{2+}$, Li$^{3+}$, and C$^{6+}$. In the FBA, the motion of the incident and scattered projectiles is treated as a plane wave in both the initial and final state. All single electron bound states are given by hydrogenic wave functions that are analytically known.

A primary advantage of the FBA is that the majority of the calculation can be performed analytically. This simplifies the analysis of a comparison between the 3-Body and 4-Body models and results in a direct link between the observed differences in the FDCS and their sources. For the analytical calculations it is useful to keep in mind the 1s hydrogen-like atom momentum space wave function for a with nuclear charge Z

$$\phi^{FT}(\vec{p}) = \frac{1}{(2\pi)^{3/2}} \int e^{i\vec{p}\cdot\vec{x}} \phi(\vec{x}) d\vec{x} = \frac{4Z^{5/2}}{\pi\sqrt{2}(Z^2+p^2)^2}. \tag{4}$$

Prior to any analytical calculations, the 4-Body model requires a 9-dimensional spatial integral and the 3-Body calculation requires a 6-dimensional integral. Numerical integration of these integrals is not feasible [29], however use of Eq. (4) allows them to be reduced to 3-dimensional integrals which we perform with standard Gaussian quadrature techniques.

## 2.1 Active Electron 4-Body Model

In the 4-Body model, all constituent particles are explicitly included in the calculation with the transition matrix given by

$$T_{fi}^{4B} = \langle \chi_f(\vec{R}_f)\phi_c(\vec{u})\psi(\vec{r})|V_i^{4B}|\chi_i(\vec{R}_i)\Phi(\vec{r},\vec{s})\rangle, \tag{5}$$

where $\chi_{i,f}$ is the incident (scattered) projectile plane wave with center of mass momentum $\vec{k}_i(\vec{k}_f)$, $\Phi(\vec{r},\vec{s})$ is the target helium atom wave function, $\phi_c(\vec{u})$ is the bound electron wave function for the captured electron, and $\psi(\vec{r})$ is the bound state of the residual He$^+$ ion. We note that while the target electrons have been labeled here for clarity, their indistinguishability has been properly included by symmetrizing the total final state wave function with respect to the two electrons. The perturbation is given by the Coulomb interaction between the projectile and constituent particles of the helium atom

$$V_i^{4B} = \frac{Z_p Z_\alpha}{r_1} + \frac{Z_p Z_e}{r_{12}} + \frac{Z_p Z_e}{r_{13}}, \tag{6}$$

where $\vec{r}_1, \vec{r}_2,$ and $\vec{r}_3$ are the lab frame position vectors of the projectile and two atomic electrons respectively, and $Z_p, Z_\alpha$, and $Z_e$ are the charges of the projectile, target nucleus, and electron respectively. The Jacobi coordinates are related to the lab frame coordinates by

$$\vec{R}_i = \vec{r}_1 - \frac{m_e(\vec{r}_2 + \vec{r}_3)}{2m_e + m_\alpha} \tag{7}$$

$$\vec{R}_f = \frac{(m_e + m_p)\vec{r}_1}{m_e + m_p} + \frac{m_e \vec{r}_2}{m_e + m_p} - \frac{m_e \vec{r}_3}{m_e + m_\alpha} \tag{8}$$

$$\vec{s} = \vec{r}_2 - \frac{m_e \vec{r}_3}{m_e + m_\alpha} \tag{9}$$

$$\vec{r} = \vec{r}_3 \tag{10}$$

$$\vec{u} = \vec{r}_{21}. \tag{11}$$

We use a one parameter variational wave function for the target ground state helium atom given by

$$\Phi(\vec{r}, \vec{s}) = \frac{\alpha^3}{\pi} e^{-\alpha r} e^{-\alpha s} \tag{12}$$

with $\alpha = \frac{27}{16}$. The bound state wave function for the residual He$^+$ ion is simply a 1s hydrogenic wave function for charge $\beta = 2$ given by

$$\psi(\vec{r}) = \frac{\beta^{\frac{3}{2}}}{\sqrt{\pi}} e^{-\beta r}. \tag{13}$$

### 2.2 Frozen Core 3-Body Model

In the 3-Body model, the inactive electron and target nucleus are considered as a single frozen core, with charge $Z_{He^+} = 1$. The transition matrix is then given by

$$T_{fi}^{3B} = \langle \chi_f(\vec{R}_f) \phi_p(\vec{u}) | V_i^{3B} | \chi_i(\vec{R}_i) \psi(\vec{s}) \rangle, \tag{14}$$

where the projectile plane waves are the same as in the 4-Body model, as is the projectile bound state. The 3-Body perturbation is

$$V_i^{3B} = \frac{Z_p Z_{He^+}}{r_1} + \frac{Z_p Z_e}{r_{12}} \tag{15}$$

and the target atom wave function is now a single electron bound state given by Eq. (13) with $\beta = \alpha = \frac{27}{16}$. The lab frame coordinates remain the same as in the 4-Body model, with the exception that the inactive electron is not present and therefore there is no $\vec{r}_3$. The Jacobi coordinates for the 3-Body model can be found by setting $\vec{r}_3 = 0$ in Eqs (7-11).

There are two primary differences between the 3-Body and 4-Body models, both of which originate from the frozen core approximation. First, in the 4-Body model the inactive electron in the target atom changes state from the ground state of the He(1s$^2$) atom to the ground state of a hydrogen-like atom with nuclear charge $\beta = 2$. This electron is completely neglected in the 3-Body model and is therefore considered not to change state. Second, the perturbation potential is different for the 3-Body and 4-Body models. In particular, the 4-Body perturbation contains a sum of three terms, while the 3-Body perturbation is only two terms. Past work on the analysis of 3-Body and 4-Body models for single ionization has shown that the perturbation potential alters the shape of the FDCS [25,26]. Additionally, FBA models are known to predict a deep minimum in the FDCS caused by a cancellation of terms in the perturbation [7,30–32]. Therefore, we expect that the form of the perturbation will be important in determining the 3-Body and 4-Body FDCS.

### 3. Results

A direct comparison of the FDCS from the 3-Body and 4-Body models shows the effect of the frozen core approximation.  In general, there is an order of magnitude difference between the models, indicating that the use of the frozen core approximation significantly alters the FDCS.  Recent experiments using the COLTRIMS technique have produced a significant number of highly detailed data sets [1–7].  For the single capture process, most experiments were performed with proton projectiles.  In Fig. 1, results are shown for the 3-Body and 4-Body models compared with the experiment of [6,7].  We chose a select few data sets that cover a wide range of projectile energies in which the FBA models have validity.  There are many more data sets available for comparison, but the results here demonstrate the features and trends of interest.

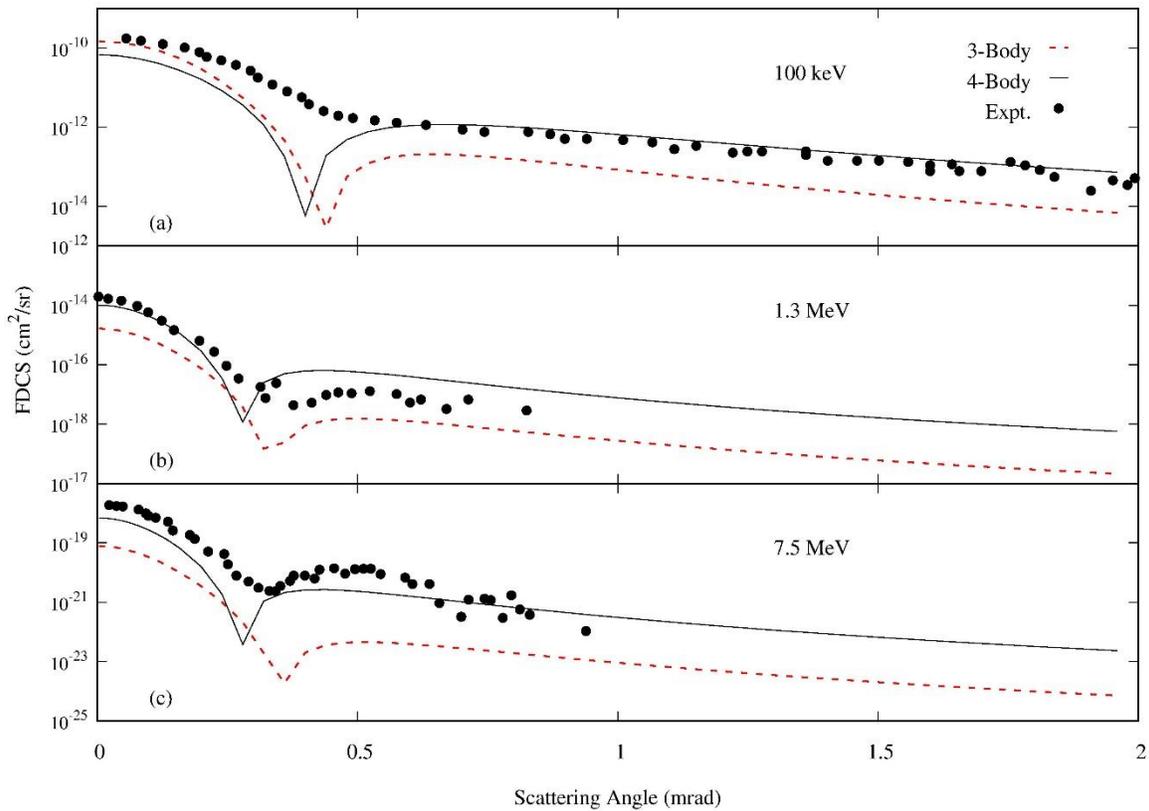

**Figure 1 Fully differential cross sections for p + He single electron capture to the ground state with the residual $He^+$ ion also in the ground state.  Experimental results are**

**from (a) [7] (b,c) [6]. Incident projectile energies are (a) 100 keV, (b) 1.3 MeV, and (c) 7.5 MeV.**

There are some important features to note in both the experimental and theoretical FDCS results. First, the experimental FDCS exhibit an elbow structure around 0.3 to 0.5 mrad where the slope of the curve changes. This elbow occurs at the boundary between small and large angle scattering and separates different physical mechanisms. At small angles, projectile-nuclear effects are less important than at large scattering angles. Physically, single capture at small angles is predominantly caused by momentum transfer to the electron, while at large scattering angles, capture occurs through momentum transfer between the nuclei [33–35].

Also, the 3-Body and 4-Body FBA models show a pronounced minimum in the FDCS that is not present in the experimental results. This minimum comes from a cancellation of terms in the perturbation [7,30–32] and is generally observed in first order perturbative models. Higher order and non-perturbative models do not have this deep minimum.

The order of magnitude difference between the 3-Body and 4-Body models is evident from Fig. 1, with the 4-Body model FDCS generally larger than the 3-Body model FDCS. The one exception to this is for a projectile energy of 100 keV at small scattering angles, where the 4-Body FDCS is slightly smaller than the 3-Body FDCS. This similarity of the models at low projectile energy and small scattering angles has also been observed in the $He^+$ + He single capture collision system [27] and can be attributed to nuclear screening being less important in grazing collisions. In general, the 4-Body model is in better agreement with experiment than the 3-Body model, although neither is expected to exactly predict experiment given the simplicity of the models. Because the 4-Body model better predicts experiment, one can conclude that the inactive electron plays an important role in the single capture process.

To further explore exactly how the inactive electron affects the FDCS, we perform a more comprehensive comparison and analysis of the 3-Body and 4-Body models. Figures 2 and 3 contain FDCS for four projectiles ($H^+$, $He^{2+}$, $Li^{3+}$, and $C^{6+}$) and 3 projectile velocities (100 keV/amu, 1 MeV/amu, 10 MeV/amu). Some general trends can be observed. As projectile velocity increases, the magnitude of the FDCS decreases, as is expected from the total cross sections, which decrease with increasing projectile energy [15]. Also, as projectile charge and mass increase, the FDCS increase in magnitude, making capture more likely for heavier and more highly charged ions. The deep minimum in the FDCS that is characteristic of FBA models occurs at different scattering angles for the 3-Body and 4-Body models. This is expected since the minimum is caused by a cancellation of terms in the perturbation potential, which is different for the two models. The location of the minimum moves to smaller scattering angles as projectile charge and mass increase. However, because projectile charge factors out of the perturbation and results in an overall multiplicative constant, the shift in minimum location as the projectile changes is due to the changing mass of the projectile.

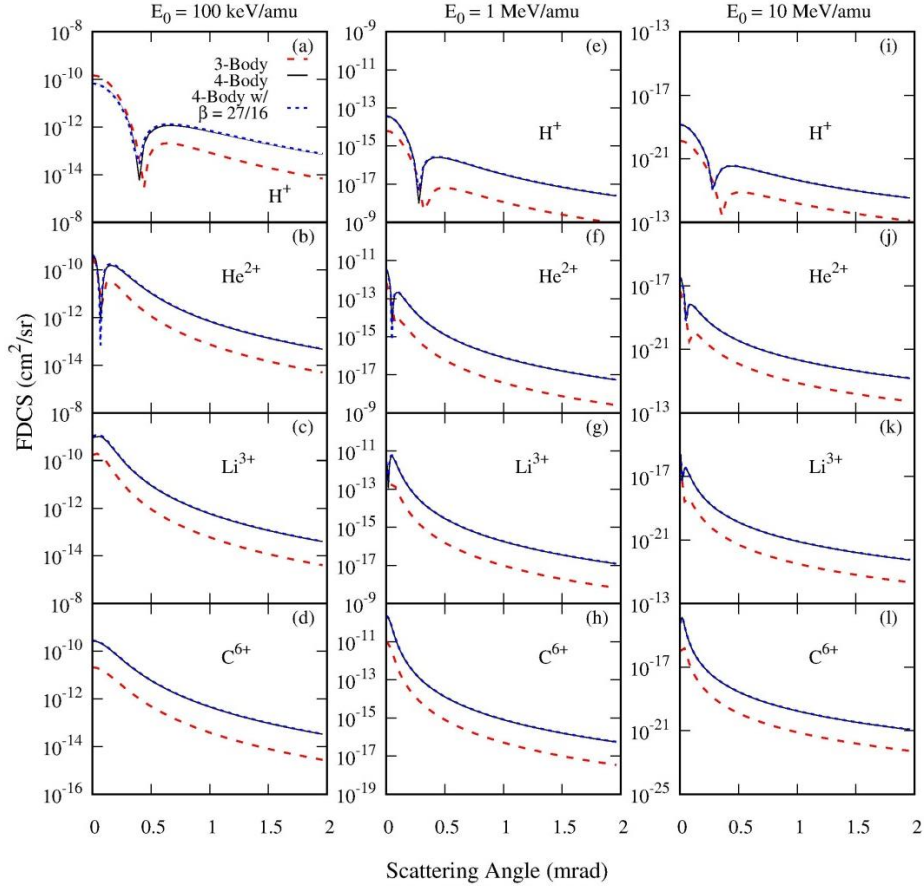

**Figure 2** Fully differential cross sections for single electron capture to the ground state of the projectile with the residual $He^+$ ion also in the ground state. The columns contain results for incident projectile energies of 100 keV/amu (column 1), 1 MeV/amu (column 2), and 10 MeV/amu (column 3). Results are present for projectiles $H^+$ (row 1), $He^{2+}$ (row 2), $Li^{3+}$ (row 3), and $C^{6+}$ (row 4).

Comparison of the 3-Body and 4-Body models shows that for all projectiles and energies, the 4-Body model is generally one order of magnitude larger than the 3-Body model. The FDCS are most similar at small scattering angles, which is the regime in which nuclear-nuclear interactions are lease important. Because the primary difference between the models is the treatment of target nuclear screening, it is understandable that the models would be most similar at small scattering angles where screening is less important. As scattering angle increases, nuclear-nuclear interactions become more important and the difference in screening treatments between the two models becomes more apparent. As projectile energy increases, the 3-Body and 4-Body models become more similar, indicating that projectile screening becomes less important for faster projectiles.

The treatment of the inactive electron in the 4-Body model differs in two fundamental ways from that of the 4-Body model. First, the inactive electron in the 4-Body model changes state from the ground state of a He($1s^2$) atom to the ground state of a He$^+$ ion. Second, the Coulomb interaction in the perturbation potential of the 4-Body model contains three terms, whereas in the 3-Body model, it only contains two terms. The 4-Body model perturbation contains the Coulomb interaction of the projectile with each constituent particle of the He$^+$ ion core, but the 3-Body model perturbation contains only an interaction for the core as a single point particle. Both of these differences between the 3-Body and 4-Body models can be individually explored through modifications to the 4-Body model.

We first examine the effect of the inactive electron changing state from its initial atomic state to its final ionic state in the 4-Body model. With the one parameter wave function used here for the helium atom, the inactive electron is effectively in a hydrogenic orbital for a nuclear charge of 27/16. After the collision, the electron has transitioned to a hydrogenic orbital for nuclear charge 2. Classically, this corresponds to the electron transitioning to a smaller orbital radius. In the 3-Body model, the inactive electron is assumed not to change state and therefore it remains in its initial bound state. The inactive electron in the 3-Body model corresponds to a 4-Body model in which the final state inactive electron is in the same hydrogenic orbital as it was in the initial state.

To study the effect of the inactive electron's change of state, we perform a 4-Body calculation in which the final state He$^+$ ion wave function is hydrogenic with nuclear charge

27/16 (the same as the initial state orbital). All other features of the 4-Body model are unaltered. Results of the 3-Body, 4-Body, and 4-Body with $\beta = 27/16$ models are shown in Fig. 2. For all projectiles and projectile velocities, the final state He$^+$ ion's nuclear charge has no discernable effect on the FDCS. Closer examination of the two calculations reveals that they differ by at most 10%, with the results for $\beta = \frac{27}{16}$ being larger than those for $\beta = 2$. Thus, we conclude that the change of state of the inactive electron is not the primary cause of the discrepancies between the 3-Body and 4-Body FDCS. This is consistent with our previous work for the single ionization process, in which the change of state of the inactive electron had a negligible effect on the FDCS [25,26].

The other difference between the 3-Body and 4-Body models is the perturbation potential used in the calculations. In the 3-Body model, the frozen core approximation assumes that the Coulomb interaction between the projectile and the He$^+$ core consists of only one term in which the core is modeled as a point particle with charge +1. To test the effect of this approximation, we perform a calculation using the 4-Body model, but replace the 4-Body perturbation of Eq. (6) with the 3-Body perturbation of Eq. (15). Results are shown in Fig. 3, and it is apparent that the choice of perturbation significantly affects the magnitude of the FDCS. In particular, the FDCS using the 4-Body model with the 3-Body perturbation are generally smaller than those of the complete 4-Body calculation, although not as small as the 3-Body model. For all projectile energies, the FDCS calculated using the 4-Body model with the 3-Body perturbation fall between the 3-Body and 4-Body models. This indicates that full target nuclear screening in the 3-Body model results in lower capture cross sections and points to the important role of nuclear-nuclear interactions in the capture process. Clearly the model chosen for the projectile-target interaction significantly alters the FDCS much more than the inactive electron changing state.

We can then conclude that the concept of the frozen core approximation, in which the inactive electron does not change state, is valid for single electron capture, but that care must be used in modeling the projectile-target interaction. These results are consistent with our past studies of the frozen core approximation and nuclear screening [25–27].

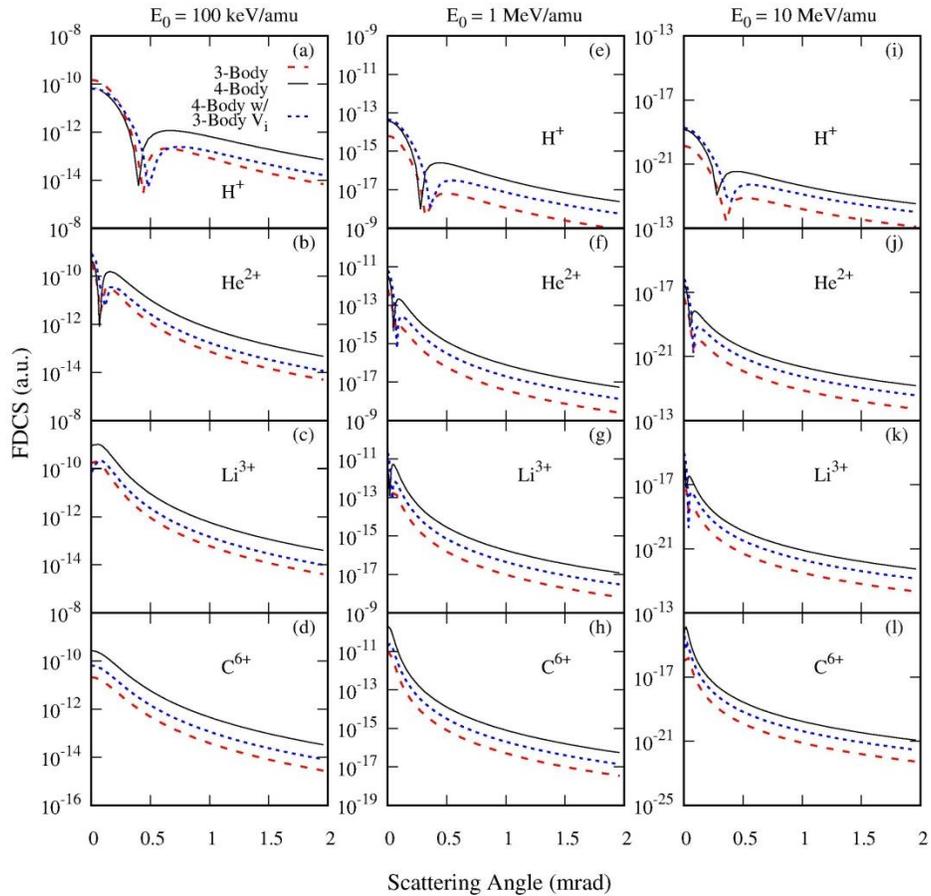

**Figure 3** Same as Fig. 2, but results are included for the 4-Body model using the 3-Body perturbation of Eq. (15), as described in the text.

In summary, we used the first Born approximation in 3-Body and 4-Body models of single electron capture to explore the effects of the frozen core approximation. Application of the models to different projectiles and projectile velocities revealed an overall magnitude difference between the models of about a factor of 10, with the 4-Body model predicting larger FDCS. The difference in magnitude between the models was traced primarily to the projectile-target Coulomb interaction used in the perturbation. The inactive electron's change of state from an atomic ground state to an ionic ground state had almost no effect on the FDCS. All of these models are consistent with prior studies of the frozen core approximation and point to the need to carefully consider how nuclear screening is included in models.

## Acknowledgements

We gratefully acknowledge the support of the NSF under Grant Nos. PHY-1505217 and PHY-1838550.

References

[1] J. Ullrich, R. Moshammer, A. Dorn, R. D. rner, L. P. H. Schmidt, and H. S.-B. cking, Rep. Prog. Phys. **66**, 1463 (2003).
[2] H.-K. Kim, M. S. Schöffler, S. Houamer, O. Chuluunbaatar, J. N. Titze, L. P. H. Schmidt, T. Jahnke, H. Schmidt-Böcking, A. Galstyan, Y. V. Popov, and R. Dörner, Phys. Rev. A **85**, (2012).
[3] V. Mergel, R. Dörner, K. Khayyat, M. Achler, T. Weber, O. Jagutzki, H. J. Lüdde, C. L. Cocke, and H. Schmidt-Böcking, Phys. Rev. Lett. **86**, 2257 (2001).
[4] M. Zapukhlyak, T. Kirchner, A. Hasan, B. Tooke, and M. Schulz, Phys. Rev. A **77**, 012720 (2008).
[5] D. Fischer, K. Støchkel, H. Cederquist, H. Zettergren, P. Reinhed, R. Schuch, A. Källberg, A. Simonsson, and H. T. Schmidt, Phys. Rev. A **73**, 052713 (2006).
[6] D. Fischer, M. Gudmundsson, Z. Berényi, N. Haag, H. A. B. Johansson, D. Misra, P. Reinhed, A. Källberg, A. Simonsson, K. Støchkel, H. Cederquist, and H. T. Schmidt, Phys. Rev. A **81**, 012714 (2010).
[7] M. S. Schöffler, J. Titze, L. P. H. Schmidt, T. Jahnke, N. Neumann, O. Jagutzki, H. Schmidt-Böcking, R. Dörner, and I. Mančev, Phys. Rev. A **79**, 064701 (2009).
[8] M. Zapukhlyak and T. Kirchner, Phys. Rev. A **80**, 062705 (2009).
[9] P. R. Simony, J. H. McGuire, and J. Eichler, Phys. Rev. A **26**, 1337 (1982).
[10] D. L. Guo, X. Ma, R. T. Zhang, S. F. Zhang, X. L. Zhu, W. T. Feng, Y. Gao, B. Hai, M. Zhang, H. B. Wang, and Z. K. Huang, Phys. Rev. A **95**, 012707 (2017).
[11] J. W. Gao, Y. Wu, J. G. Wang, N. Sisourat, and A. Dubois, Phys. Rev. A **97**, 052709 (2018).
[12] E. G. Adivi and M. A. Bolorizadeh, J. Phys. B At. Mol. Opt. Phys. **37**, 3321 (2004).
[13] H. A. Slim, E. L. Heck, B. H. Bransden, and D. R. Flower, J. Phys. B At. Mol. Opt. Phys. **24**, 2353 (1991).
[14] E. Ghanbari-Adivi and H. Ghavaminia, Phys. Scr. **89**, 105402 (2014).
[15] S. Halder, A. Mondal, S. Samaddar, C. R. Mandal, and M. Purkait, Phys. Rev. A **96**, 032717 (2017).

[16]   D. Belkić and I. Mančev, Phys. Scr. **45**, 35 (1992).
[17]   D. Belkić and I. Mančev, Phys. Scr. **47**, 18 (1993).
[18]   D. Belkić, Phys. Rev. A **47**, 189 (1993).
[19]   D. Belkic, J. Phys. B At. Mol. Opt. Phys. **26**, 497 (1993).
[20]   A. L. Harris and D. H. Madison, Phys. Rev. A **90**, 022701 (2014).
[21]   D. Belkić, I. Mančev, and J. Hanssen, Rev. Mod. Phys. **80**, 249 (2008).
[22]   I. Mancev, V. Mergel, and L. Schmidt, J. Phys. B At. Mol. Opt. Phys. **36**, 2733 (2003).
[23]   R. Samanta, M. Purkait, and C. R. Mandal, Phys. Rev. A **83**, 032706 (2011).
[24]   S. Samaddar, S. Halder, A. Mondal, C. R. Mandal, M. Purkait, and T. K. Das, J. Phys. B At. Mol. Opt. Phys. **50**, 065202 (2017).
[25]   A. L. Harris and K. Morrison, J. Phys. B At. Mol. Opt. Phys. **46**, 145202 (2013).
[26]   A. L. Harris, J. Phys. B At. Mol. Opt. Phys. **48**, 115203 (2015).
[27]   A. Harris and A. Plumadore, J. Phys. B At. Mol. Opt. Phys. (2019).
[28]   M. R. C. McDowell and J. P. Coleman, *Introduction to the Theory of Ion-Atom Collisions* (Elsevier Science Publishing Co Inc.,U.S., 1970).
[29]   A. L. Harris and D. H. Madison, Phys. Rev. A **90**, (2014).
[30]   E. Ghanbari-Adivi and H. Ghavaminia, Chin. Phys. B **24**, 033401 (2015).
[31]   D. Belkic and A. Salin, J. Phys. B At. Mol. Phys. **11**, 3905 (1978).
[32]   K. Omidvar, Phys. Rev. A **12**, 911 (1975).
[33]   V. Mergel, R. Dörner, J. Ullrich, O. Jagutzki, S. Lencinas, S. Nüttgens, L. Spielberger, M. Unverzagt, C. L. Cocke, R. E. Olson, M. Schulz, U. Buck, E. Zanger, W. Theisinger, M. Isser, S. Geis, and H. Schmidt-Böcking, Phys. Rev. Lett. **74**, 2200 (1995).
[34]   E. Y. Kamber, C. L. Cocke, S. Cheng, and S. L. Varghese, Phys. Rev. Lett. **60**, 2026 (1988).
[35]   R. Dörner, J. Ullrich, H. Schmidt-Böcking, and R. E. Olson, Phys. Rev. Lett. **63**, 147 (1989).